%% file: main.tex
\documentclass[%
 amsmath,amssymb,
 aps, twocolumn]{revtex4-1}
\usepackage{CJK}
\usepackage{graphicx}
\usepackage{dcolumn}
\usepackage{bm}
\usepackage{amsmath}
\usepackage{physics}

\usepackage{scalerel} 
\usepackage{titlesec}
\usepackage{float}
\usepackage{amssymb}
\usepackage{standalone}
\usepackage{siunitx}
\usepackage{tabularx}
\usepackage{xcolor}

\titlespacing*{\section}
{0pt}{2.5ex plus 1ex minus .2ex}{2.3ex plus .2ex}
\titlespacing*{\subsection}
{0pt}{2.5ex plus 1ex minus .2ex}{2.3ex plus .2ex}

\def\app#1#2{%
  \mathrel{%
    \setbox0=\hbox{$#1\sim$}%
    \setbox2=\hbox{%
      \rlap{\hbox{$#1\propto$}}%
      \lower1.1\ht0\box0%
    }%
    \raise0.25\ht2\box2%
  }%
}

\begin{document}

\begin{CJK*}{GB}{}

\title[Characterisation of spatial charge sensitivity in a multi-mode superconducting qubit]{Characterisation of spatial charge sensitivity in a multi-mode superconducting qubit}

\author{J. Wills}
\author{G. Campanaro}
\author{S. Cao}
\author{S. D. Fasciati}
\author{P. J. Leek}
\author{B. Vlastakis}
\thanks{Now at Oxford Quantum Circuits, Thames Valley Science Park, Reading RG2 9LH; bvlastakis@oxfordquantumcircuits.com}

\affiliation{Clarendon Laboratory, Department of Physics, University of Oxford, Oxford OX1 3PU, United Kingdom}

\date{\today}
\begin{abstract}
Understanding and suppressing sources of decoherence is a leading challenge in building practical quantum computers. In superconducting qubits, low frequency charge noise is a well-known decoherence mechanism that is effectively suppressed in the transmon qubit. Devices with multiple charge-sensitive modes can exhibit more complex behaviours, which can be exploited to study charge fluctuations in superconducting qubits. Here we characterise charge-sensitivity in a superconducting qubit with two transmon-like modes, each of which is sensitive to multiple charge-parity configurations and charge-offset biases. Using Ramsey interferometry, we observe sensitivity to four charge-parity configurations and track two independent charge-offset drifts over hour timescales. We provide a predictive theory for charge sensitivity in such multi-mode qubits which agrees with our results. Finally, we demonstrate the utility of a multi-mode qubit as a charge detector by spatially tracking local-charge drift.
\end{abstract}
\maketitle
\end{CJK*}


\section{INTRODUCTION}
\input{introduction.tex}

\section{A TWO-MODE COAXIAL TRANSMON}
\input{theory.tex}
\section{METHODS}
\input{methods.tex}

\section{RESULTS}
\input{results.tex}
\section{CONCLUSION}
\input{conclusion.tex}

\begin{acknowledgments}
This work has received funding from the United Kingdom Engineering and Physical Sciences Research Council under Grants No. EP/J013501/1, EP/M013243/1, EP/N015118/1, EP/T001062/1 and from Oxford Quantum Circuits Limited. S.D.F. acknowledges support from the Swiss Study Foundation and the Bakala Foundation. B.V. acknowledges support from an EU Marie Sklodowska-Curie fellowship.
\end{acknowledgments}
\section{APPENDIX}
\input{appendix.tex}

\newpage

\bibliographystyle{apsrev4-1}

\bibliography{References}


\end{document}

%% file: introduction.tex
Realising practical quantum computation requires the low-error operation of many fully controlled quantum bits with individual state readout and initialisation \cite{DiVincenzo2000}. Superconducting circuits are well established as a potential platform for quantum computation \cite{Devoret2013, Arute_2019}. One superconducting qubit variant, the transmon \cite{Koch2007}, has found enduring success due to its resilience to decoherence and simplicity in design. However the development of alternatives is an active area of research \cite{Manucharyan2009, Kou2017, Groszkowski2018, Gyenis2021}. One particular issue for the transmon is its fixed and sizeable coupling to other qubits and circuit modes due to its large dipole moment, which can cause unwanted crosstalk and non-negligible computational errors \cite{Sundaresan2020}. This issue can be addressed by using tuneable couplers \cite{Roy2017, Srinivasan2011}, level-structure engineering \cite{Richer2017, Noguchi2020}, or auxiliary circuit modes \cite{Smith2020}.

Here we investigate the multi-mode transmon qubit. Such qubits allow for strongly coupled, and far detuned modes, that can assist in preventing cross-talk \cite{Finck2021} and limit contributions to decay due to the Purcell effect \cite{Gambetta2011}. The energy level structure can be engineered to allow for generous resonance conditions and fast entangling operations \cite{Roy2018}. In addition, this can create mechanisms for photon-dephasing suppression \cite{Richer2017, Zhang2017} or tuneable coupling \cite{Gambetta2011}, making multi-mode qubits of potential use in superconducting quantum processors.

Since the modes of these multi-mode systems are transmon-like, they are susceptible to the same decoherence mechanisms, and can exhibit more complex sensitivities to noise \cite{Pashkin2003}. One such decoherence mechanism that affects superconducting circuits is low frequency, $1/f$-like charge noise \cite{Gustafsson2013}. This has been well characterised and studied in single-mode transmons, which are typically designed to operate in a charge-insensitive regime \cite{Schreier2008}. 
Alternatively, devices can be deliberately designed to be sensitive to charge noise, and operate as detectors of it, in order to better understand and characterise this source of decoherence \cite{Christensen2019, Wilen2021, tennant2021}. The  origins of charge noise in superconducting circuits are theorised to be surface drifts, patch potentials, voltage fluctuations in control electronics, or the effects of the absorption of cosmic radiation within substrates \cite{Christensen2019, Wilen2021, Martinis2021, tennant2021}.

\begin{figure}[ht]
\includegraphics[width=0.45\textwidth]{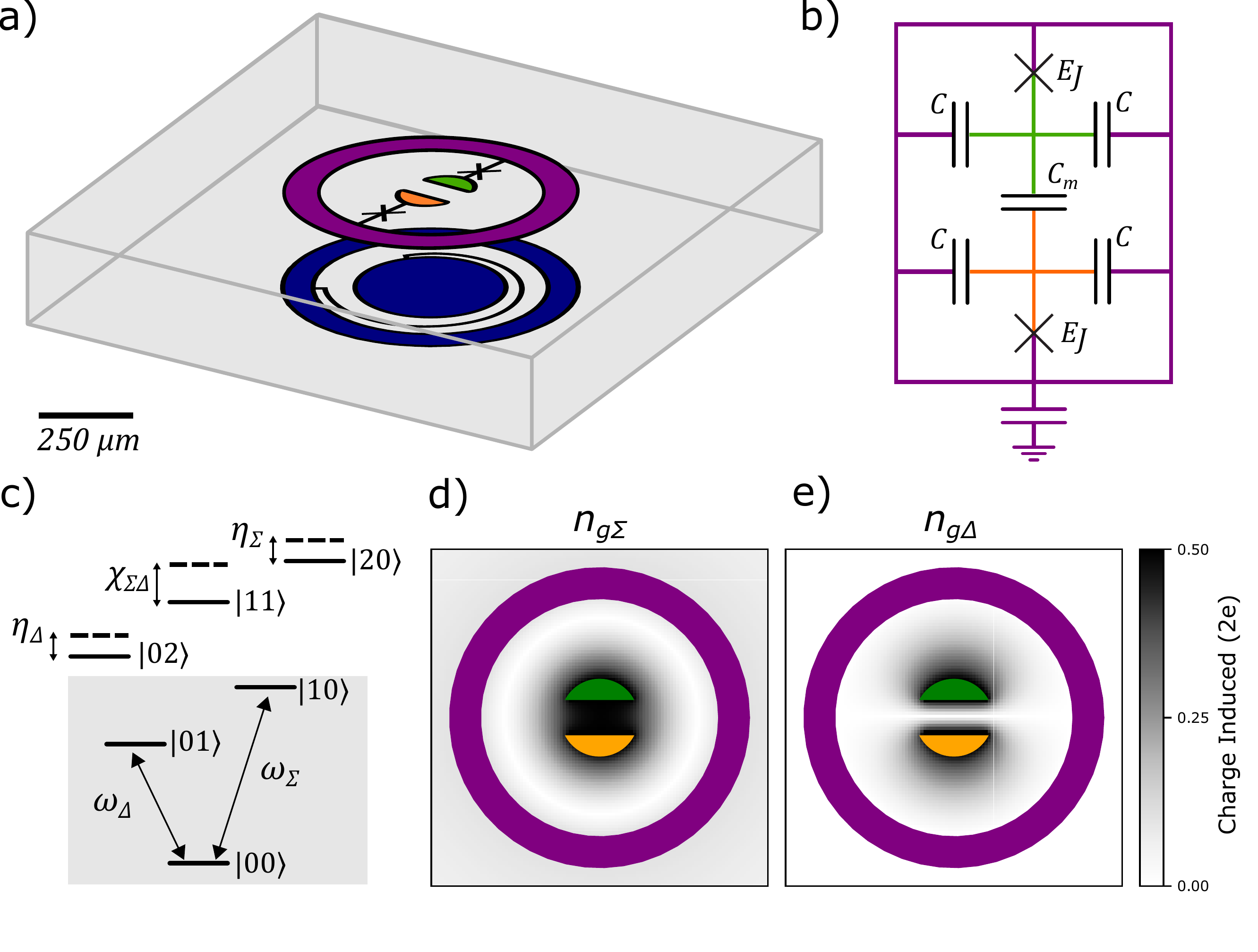}
\caption{Device Description. (a) A simplified schematic of the device shows a coaxial pad geometry with a split inner conductor resulting in three superconducting islands. Two near-identical Josephson junctions with phase $\varphi_1$, $\varphi_2$ are used resulting in two non-degenerate modes ($\Delta, \Sigma$). A lumped element LC resonator (blue) on the opposing side of substrate allows for dispersive readout of modes. (b) An equivalent circuit of the two-mode coaxial transmon. The two inner superconducting islands are connected by a coupling capacitance ($C_m$), and connected to the outer superconducting island by Josephson junction ($E_J$) and capacitance ($2C$). (c) Energy level diagram of the two-mode coaxial transmon illustrating anharmonicities ($\eta_{\Sigma,\Delta}$), and state-dependent shift ($\chi_{\Sigma \Delta}$). States labelled as $\ket{nm}$ where $n (m)$ corresponds to the number of excitations in the $\Sigma$-mode ($\Delta$-mode). The V-shaped qutrit subspace is highlighted in grey. (d) (e) Simulation of charge bias induced by a point charge (1e) in the summation ($n_{g\Sigma}$, left), and difference ($n_{g\Delta}$, right) charge configurations on the two inner islands.}
\label{figure:1}
\end{figure}

In this paper, we investigate the effect of charge noise in a two-mode transmon design, depicted in Fig.~\ref{figure:1}. We present a description of the mode structure and device operation, as well as a predictive theory for charge sensitivity in such devices. We observe sensitivity to the four charge-parity configurations that arise as a result of the two degrees of freedom in a charge-sensitive two-mode transmon qubit. We also demonstrate effective suppression of charge sensitivity in an alternative design two-mode device with high $E_J/E_C$ ratio, in accordance with our predictive theory. Finally, using Ramsey interferometry we track charge-offset fluctuations showing a proof-of-concept detector for spatial drifts in charge noise over $100~\rm{\mu m}$ length scales.

Understanding decoherence mechanisms, such as charge noise, is vital to the application of superconducting qubits in a low-error quantum processor. In turn, a charge-sensitive multi-mode transmon could prove a useful tool for identifying the origins of such charge fluctuations in high-coherence quantum devices.

%% file: theory.tex
\begin{figure*}[ht]
\includegraphics[width=1.\textwidth]{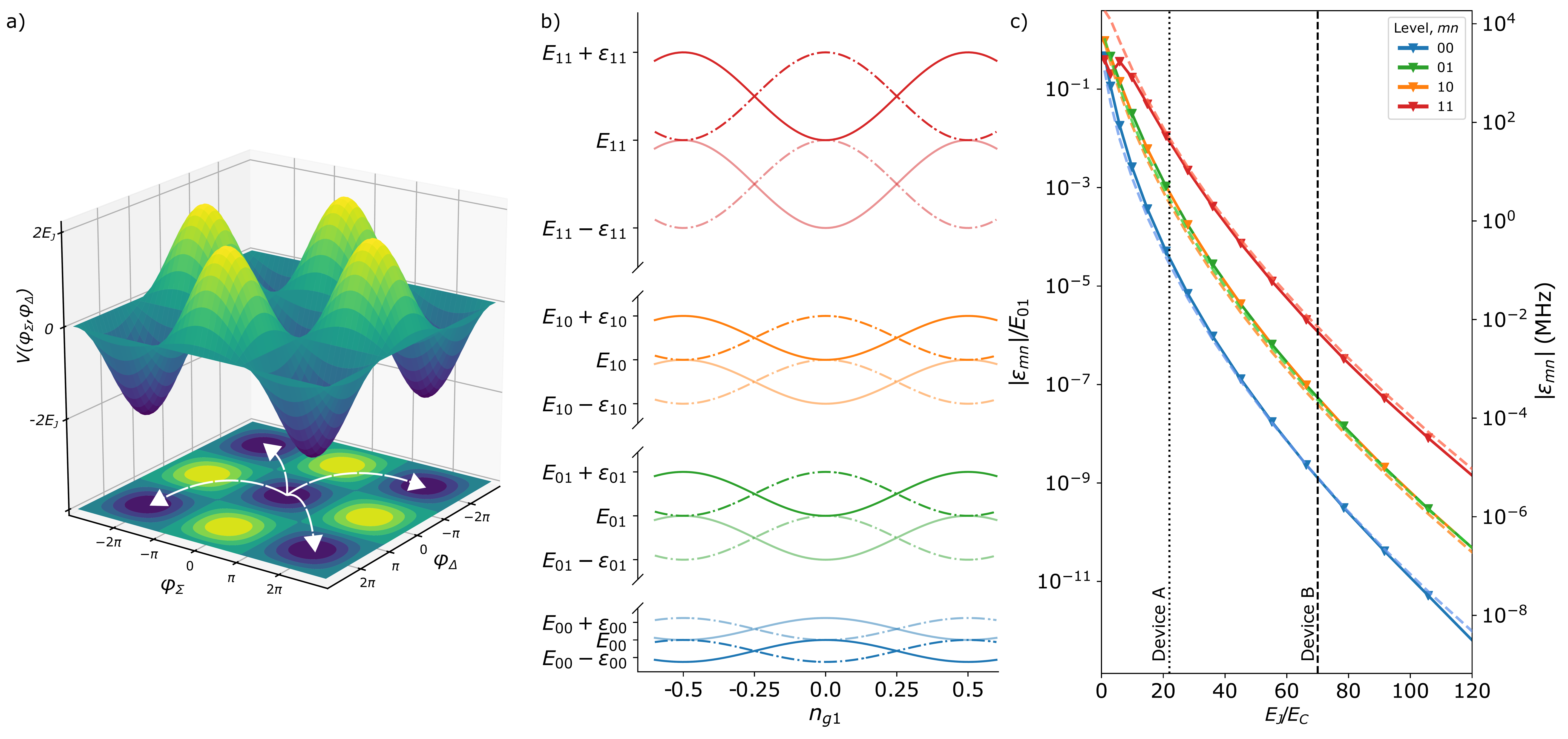}
\caption{Charge dispersion and parity configurations in a multimode system. (a) The $V(\varphi_\Sigma,\varphi_\Delta ) = - 2E_J \cos{\frac{\varphi_\Sigma}{2}} \cos{\frac{\varphi_\Delta}{2}}$  periodic potential energy landscape of the device, where local minima are  $2\pi$ periodic along the diagonals. White dashed arrows indicate the four possible tunnelling routes to nearest neighbouring lattice sites. The probability of tunnelling to a neighbouring well in each of the four directions is equal. (b) Dispersion of eigenenergies of the four lowest energy levels of the qubit Hamiltonian as a function of gate charge offset $n_{g1}$, with constant gate charge offset $n_{g2}$, showing the four possible parity configurations (OO, EO, OE, EE). (c) Comparison of numerical (solid) and tight binding model (dashed) calculations of charge dispersion for the lowest four energy levels of the qubit Hamiltonian, as a function of the ratio $E_J/E_C$. The right vertical scale gives the charge dispersion in MHz for a lowest transition frequency of 5 GHz. Vertical dotted (dashed) line shows the regime of the charge sensitive Device A, $E_J/E_C$ = 22 (charge suppressed Device B, $E_J/E_C$ = 70).}
\label{figure:2}
\end{figure*}

A transmon qubit is a simple superconducting circuit consisting of a capacitance in parallel with a Josephson junction, and is insensitive to charge fluctuations across the capacitor in the regime of large Josephson-to-charging energy ratio, $E_J/E_C$ \cite{Koch2007}. Here we work with a circuit with two transmon-like modes, built from three superconducting islands and two Josephson junctions, depicted in Fig.~\ref{figure:1}(b).

The two-mode circuit has the Hamiltonian:
\begin{equation}
\begin{aligned}
    \hat{H} = &4E_{C}(\hat{n}_1 - n_{g1})^2 + 4E_{C}(\hat{n}_2 - n_{g2})^2 \\
    & + 4E_{p} (\hat{n}_1 - n_{g1})(\hat{n}_2 - n_{g2}) \\
    & - E_J \cos{\hat{\varphi}_1} - E_J \cos{\hat{\varphi}_2}   
\end{aligned}
\label{equation:1}
\end{equation}
where the charging energy $E_C = e^2 C^* / 2(C^{*2} - C_m^2)$, with $C^* = C(\frac{C+2C_m}{C+C_m})$ the total shunting capacitance of each island. $E_J$ is the Josephson energy of each junction, and $E_p = e^2 C_m/({C^*}^2 - C_m^2)$ is the coupling energy between the two modes \cite{Pashkin2003}. The Hamiltonian of this system is identical to that of two resonantly-coupled transmons, the eigenmodes of which are sum and difference modes with $\hat{\varphi}_\Sigma = \hat{\varphi}_1 + \hat{\varphi}_2$ and $\hat{\varphi}_\Delta = \hat{\varphi}_1 - \hat{\varphi}_2$ respectively. In this basis, the Hamiltonian of the circuit becomes:
\begin{equation}
\begin{aligned}
\hat{H}  = &8E_{C_\Sigma}(\hat{n}_\Sigma - n_{g\Sigma})^2 + 8E_{C_\Delta}(\hat{n}_\Delta - n_{g\Delta})^2 \\
     & - 2E_J \cos{\frac{\hat{\varphi}_\Sigma}{2}} \cos{\frac{\hat{\varphi}_\Delta}{2}}
\end{aligned}
\label{equation:2}
\end{equation}
where $E_{C_\Sigma} = E_C + E_p/2 = e^2 / 2(C^*-C_m)$ and $E_{C_\Delta} = E_C - E_p/2 = e^2 / 2(C^*+C_m)$ are the charging energies of the sum and difference modes respectively.

In this representation, it is clearer that the first two excited states of this system correspond to an excitation in either the in-phase ($\Sigma$-mode), or out-of-phase oscillations ($\Delta$-mode) of the two inner superconducting islands, shown in Fig.~\ref{figure:1} (a). The antisymmetric mode is lower in frequency, has an electric dipole moment and couples well to electric fields polarised in the plane of the device. The symmetric mode is higher in frequency, has an electric quadrupole moment and so couples well to radial fields and coaxial control ports. This difference in polarisation symmetry can be used for Purcell protection \cite{Gambetta2011}.

Starting from the Hamiltonian of Eqn.~\ref{equation:2}, we can perform a second quantisation by expanding the cosine potential terms to fourth order, and assuming a weakly non-linear oscillator-like behaviour. Keeping only counter-rotating terms produces:

\begin{equation}
\begin{aligned}
    \hat{H}/\hbar &= \omega_\Sigma \hat{a}^\dagger_\Sigma \hat{a}_\Sigma + 
    \omega_\Delta \hat{a}^\dagger_\Delta \hat{a}_\Delta \\
    &-\eta_\Sigma/2 *  \hat{a}^\dagger_\Sigma{}^2 \hat{a}_\Sigma{}^2- \eta_\Delta/2 *  \hat{a}^{\dagger}_\Delta{}^2 \hat{a}_\Delta{}^2 \\
    &- \chi_{\Sigma\Delta} * \hat{a}^\dagger_\Sigma \hat{a}_\Sigma \hat{a}^\dagger_\Delta \hat{a}_\Delta
\end{aligned}
\label{equation:3}
\end{equation}
where $\hat{a}_{\Sigma(\Delta)}^{(\dagger)}$ is the annihilation (creation) operator, $\omega_{\Sigma (\Delta)}$ is the frequency and $\eta_{\Sigma (\Delta)}$ is the anharmonicity of the $\Sigma$-mode ($\Delta$-mode). $\chi_{\Sigma\Delta}$ is the non-linear coupling between the modes. This Hamiltonian is derived in Appendix A.

Both modes are very strongly coupled to each other through the junctions of the device. Notably, the non-linear coupling ($\chi_{\Sigma \Delta}$) is larger than their respective anharmonicities ($\eta_\Sigma$, $\eta_\Delta$). This produces addressable transitions that make up an effective V-shaped qutrit energy diagram \cite{Dumur2015}, as shown in Fig.~\ref{figure:1} (c). The states are labelled as $\ket{nm}$, where $n (m)$ corresponds to the number of excitations in the $\Sigma$-mode ($\Delta$-mode). We have the potential to use the qutrit system for all-microwave two-qubit gates \cite{Hazra2020, Finck2021}, using one transition as a computational bit and the other to generate entanglement with other qutrits. This allows for far-detuned computational transitions for minimised single-qubit-gate crosstalk, while retaining gate speeds comparable to fixed coupled transmons. These features make the two-mode transmon a potentially useful component for an extensible quantum computing architecture. 

We build this system in a coaxial geometry with out-of-plane coupling to a lumped-element resonator for dispersive readout \cite{Rahamim2017}. The designed circuit has a relative coupling $E_p/E_C$ $= 2.5$, which is approximately two orders of magnitude larger than typical qubit-qubit couplings.

The Hamiltonian of Eqn.~\ref{equation:2} shows a dependence on gate-charge offsets $n_{g\Sigma}$ and $n_{g\Delta}$, corresponding to the sum and difference of gate-charge offsets of the two inner islands, $n_{g1}$ and $n_{g2}$. Using electrostatic simulations, we find the induced spatially dependant offset charge due to a point charge on the surface of the substrate, shown in Fig.~\ref{figure:1} (d) and (e). The sensitivity pattern differs for $n_{g\Sigma}$ and $n_{g\Delta}$ due to the symmetric/anti-symmetric behaviour of each configuration. This difference crucially allows us to spatially detect local charge fluctuations. 

We can understand the nature and behaviour of charge dispersion in Eqn.~\ref{equation:2} in a transmon regime ($E_C,E_p \ll E_J$) by using a 2D tight-binding approximation (see Appendix D). In Fig.~\ref{figure:2} (a), the potential energy landscape shows that for each energy minimum, there are four neighbouring lattice sites dependant on $n_{g1}$ and $n_{g2}$, with an equal tunnel-barrier energy between them. This will produce a dispersion relationship of the form $E(n_{g1}, n_{g2}) \sim ( \cos{n_{g1}} + \cos{n_{g2}})$.

This can be represented in terms of the sum and difference offset charges, $n_{g\Sigma}$ and $n_{g\Delta}$, as: 
\begin{equation}
E_{mn}(n_{g \Sigma}, n_{g \Delta}) \approx \overline{E_{mn}} + \frac{\epsilon_{mn}}{4}\cos{\pi n_{g\Sigma}}\cos{\pi n_{g \Delta}}
\label{equation:4}
\end{equation}
where $\epsilon_{mn}$ is the maximum measured charge dispersion for the $E_{mn}$ level, where $m$($n$) is the number of excitations in the $\Sigma$-mode ($\Delta$-mode).

One source of decoherence is sudden changes in offset charge due to tunnelling of quasiparticles across Josephson junctions from one superconducting island to another \cite{Serniak2018}. This corresponds to jumps in either $n_{g1}\rightarrow n_{g1} + 0.5$, $n_{g2}\rightarrow n_{g2} + 0.5$, otherwise denoted as jumps in charge parity Odd (O) to Even (E) or Even to Odd. This results in four different parity configurations, two for each mode in the system. Fig.~\ref{figure:2} (b) shows the energy-level dispersion with dependence on four parity configurations (OO, EO, OE, EE), with a maximum dispersion of $\epsilon_{mn}$.

In Fig.~\ref{figure:2} (c) we show numerical calculation of charge dispersion $\epsilon_{mn}$ as a function of $E_J$ and $E_C$ for $E_p = 0.4 E_C$, obtained by calculating the eigenvalues of the Hamiltonian in Eqn.~\ref{equation:1} in the charge basis. We use a semi-analytical wavefunction approach \cite{Catelani2011} to derive an analytical form of this dependence:

\begin{equation}
\begin{aligned}
\epsilon_{mn} &\approx A_0 E_J \frac{2^{2(m+n)}}{m!n!}\\
& \times \left(\frac{E_J}{E_C(1+E_p/2)}\right)^{m/2}\left(\frac{E_J}{E_C(1-E_p/2)}\right)^{n/2}\\
& \times \exp{-\left(\sqrt{\frac{2E_J}{E_C(1+E_p/2)}} + \sqrt{\frac{2E_J}{E_C(1-E_p/2)}}\,\right)}
\end{aligned}
\label{equation:5}
\end{equation}

This charge dispersion follows an exponential suppression, and in the limit $E_p \to 0$, the exponent term tends towards the value for the standard transmon. The factor $A_0$ is obtained empirically by fitting to the numerical calculation of charge dispersion, shown in Fig.~\ref{figure:2} (c).

Each mode is sensitive to both charge offsets, i.e. if there are fluctuations in $n_{g\Sigma}$, this can be observed in dispersion of the $\Delta$-mode. Importantly, this means that a single mode can detect fluctuations in both offset charges simultaneously.

%% file: methods.tex
In this report, we present two devices. One device (A), is designed for charge sensitivity, and the other (device B) is designed for charge-noise suppression. We fabricate device A (B) through electron beam lithography (and photolithography), patterning both sides of a $0.5~{\rm mm}$ sapphire (silicon) substrate. Each device is mounted inside an aluminium sample holder within a mu-metal magnetic shield, anchored to the 10 mK stage of a dilution refrigerator, operating with a standard cQED experimental setup \cite{Spring2021}. A comparison of the device designs is shown in Appendix B. 

A state-dependent resonator frequency shift $\chi_{\Sigma(\Delta) r}$ exists for each mode, allowing for simultaneous dispersive readout of the multi-mode state of the device, as defined in Appendix A. Using qubit spectroscopy, we find the transition frequencies for device A to be $\omega_\Delta / 2\pi = 5.51$ GHz, and $\omega_\Sigma / 2\pi = 6.71$ GHz, with anharmonicites of $\eta_\Delta / 2\pi = -0.38$ GHz and $\eta_\Sigma / 2\pi = -0.34$ GHz. We identify the $\ket{00} - \ket{11}$ transition at $(\omega_\Delta + \omega_\Sigma - \chi_{\Sigma \Delta})/4\pi = 5.86$ GHz, showing an inter-modal state-dependent shift of $\chi_{\Sigma \Delta} / 2\pi = 0.50$ GHz. These values are consistent with numerical solutions obtained with finite-element simulation methods and energy-participation-ratio (EPR) analysis \cite{Minev2021}. A summary of measured parameters for both devices is shown in Table~\ref{table:I}, in Appendix B.

From these parameters, we use numerical methods to estimate values of $E_J / h = 11$ GHz, $E_C / h = 0.5$ GHz, and $E_p / h = 0.2$ GHz, as defined in section II. Given these parameters, we estimate a charge dispersion of the lowest two modes to be 4 MHz and 4.1 MHz, calculated numerically. This predicted charge dispersion and $E_J/E_C$ regime is shown in Fig.~\ref{figure:2} (c) for both devices.

We next perform time-domain measurements of the energy relaxation time $T_1$ and spin-echo coherence time $T_{2E}$ of the two modes of the device. We perform 100 repeated measurements over the course of 3 hours to find  the values reported in Table~\ref{table:I}. The fact that $T_1^{\Delta}\gg T_1^{\Sigma}$ is likely to originate from the intentional difference in geometry between the two modes, and hence their coupling to the coaxial output ports, as well as the difference in detuning and coupling to the readout resonator \cite{Gambetta2011}.

\begin{figure}
\includegraphics[width=0.48\textwidth]{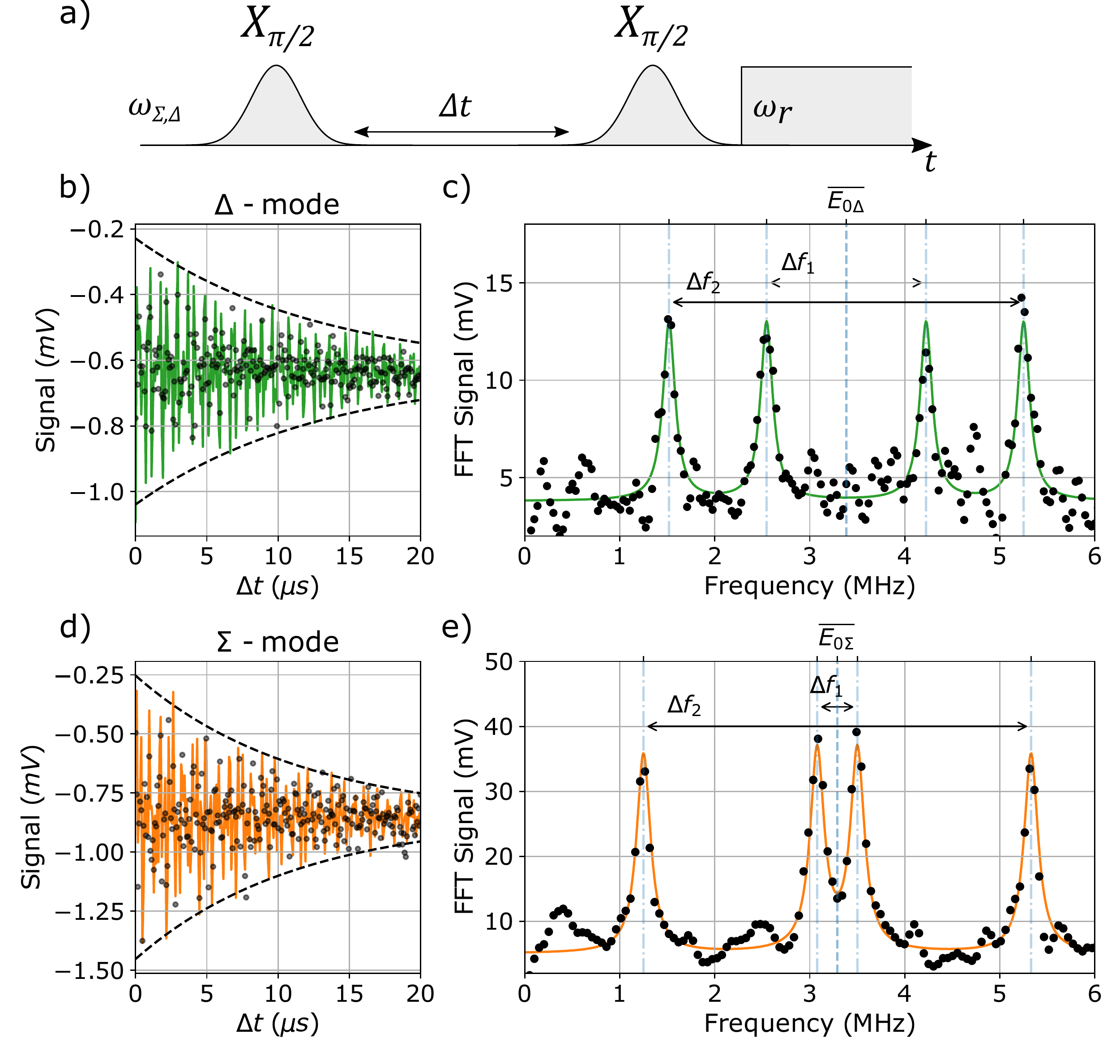}
\caption{Ramsey interferometry based measurement of charge bias configuration. Experimental measurement of charge dispersion and four parity configurations in the coaxial multi-mode qubit, with data points in black and fits in solid lines. (a) Pulse sequence for the Ramsey interferometry measurement. (b) (c) Ramsey oscillation performed on the $\Delta$-mode. The solid line in (b) shows a decaying oscillation with frequency components, and decay constant, obtained from fitting the Fourier transform of the oscillation, shown in (c). The dashed line shows an exponential decay envelope with this same fitted decay constant. The Fourier transform of the measured $\Delta$-mode Ramsey oscillations shows four frequency peaks, corresponding to the four possible parity configurations. The data is fitted to four Lorentzian peaks, symmetric about the average frequency. We define the separation of the inner peaks to be $\Delta f_1$, and the separation of the outer peaks to be $\Delta f_2$. From this separation of frequency peaks we can extract the charge bias configuration. (d) (e) Ramsey oscillation and Fourier transform of measured oscillation performed on the $\Sigma$-mode.}
\label{figure:3}
\end{figure}

We use Ramsey interferometry in order to measure energy dispersion. A mode is prepared in a superposition state using an $X_{\pi/2}$ pulse, allowed to idle for time $\Delta t$, before a second $X_{\pi/2}$ pulse is applied, as shown in the pulse sequence in Fig.~\ref{figure:3} (a). We detune the frequency of the control pulses from the average mode frequency by approximately 3.5 MHz to prevent aliasing for large frequency excursions. At each $\Delta t$, we sample 2500 times, taking approximately 100 ms to acquire. Typical transmon devices exhibit a quasiparticle tunnelling rate of $0.01~\rm{\mu s^{-1}}$ \cite{Serniak2018}, therefore we expect to average over all possible parity configurations during the measurement. As a result, we expect to observe four frequency components in a Ramsey oscillation, corresponding to the four parity configurations, shown in Fig.~\ref{figure:2} (b).

In Fig.~\ref{figure:3} we show example Ramsey oscillations measured for both the $\Delta $ - mode and $\Sigma$ - mode. Fig.~\ref{figure:3} (c) and (e) show the fast fourier transform (FFT) of the Ramsey oscillations, in which we observe four distinct frequency components. We fit the FFT data to four identical Lorenztian peaks, which are symmetric about the average frequency, $\overline{E_{0\Sigma(\Delta)}}$.  The separation of  the  two  inner  peaks  closest  to  the  symmetry  point, is labelled as $\Delta f_1$, and the separation of the two outermost peaks is labelled as $\Delta f_2$, as shown in Fig.~\ref{figure:3}. Using our tight-binding model, we find that: 

\begin{equation}
\begin{aligned}
\Delta f_1 = \frac{\epsilon_{mn}}{h} \sin{\pi n_{g\Sigma}} \sin{\pi n_{g\Delta}}, \\
\Delta f_2 = \frac{\epsilon_{mn}}{h} \cos{\pi n_{g\Sigma}} \cos{\pi n_{g\Delta}}
\end{aligned}
\end{equation}

This allows us to determine the charge configuration from the energy dispersion. Repeating this measurement, we are able to track frequency fluctuations due to correlated and anti-correlated charge noise dynamics over extended periods of time.

Note that there are several technical shortcomings of our demonstration experiment, which can be remedied in the future. Firstly, there is no gate charge control in this current architecture, which prevents us from resolving jumps or drifts larger than 0.5e for $n_{g1}$ and $n_{g2}$. This limits our ability to determine a charge noise spectral density at this time \cite{Christensen2019}, but can be alleviated with the incorporation of local control of static electric fields via gate electrodes. Secondly, the time taken to acquire each Ramsey oscillation trace limits the ability to observe changes in charge configuration faster than two minutes. This can be remedied using higher fidelity readout with a parametric amplifier, or a more efficient sampling of Ramsey delay times.

%% file: results.tex
\begin{figure}
\includegraphics[width=0.48\textwidth]{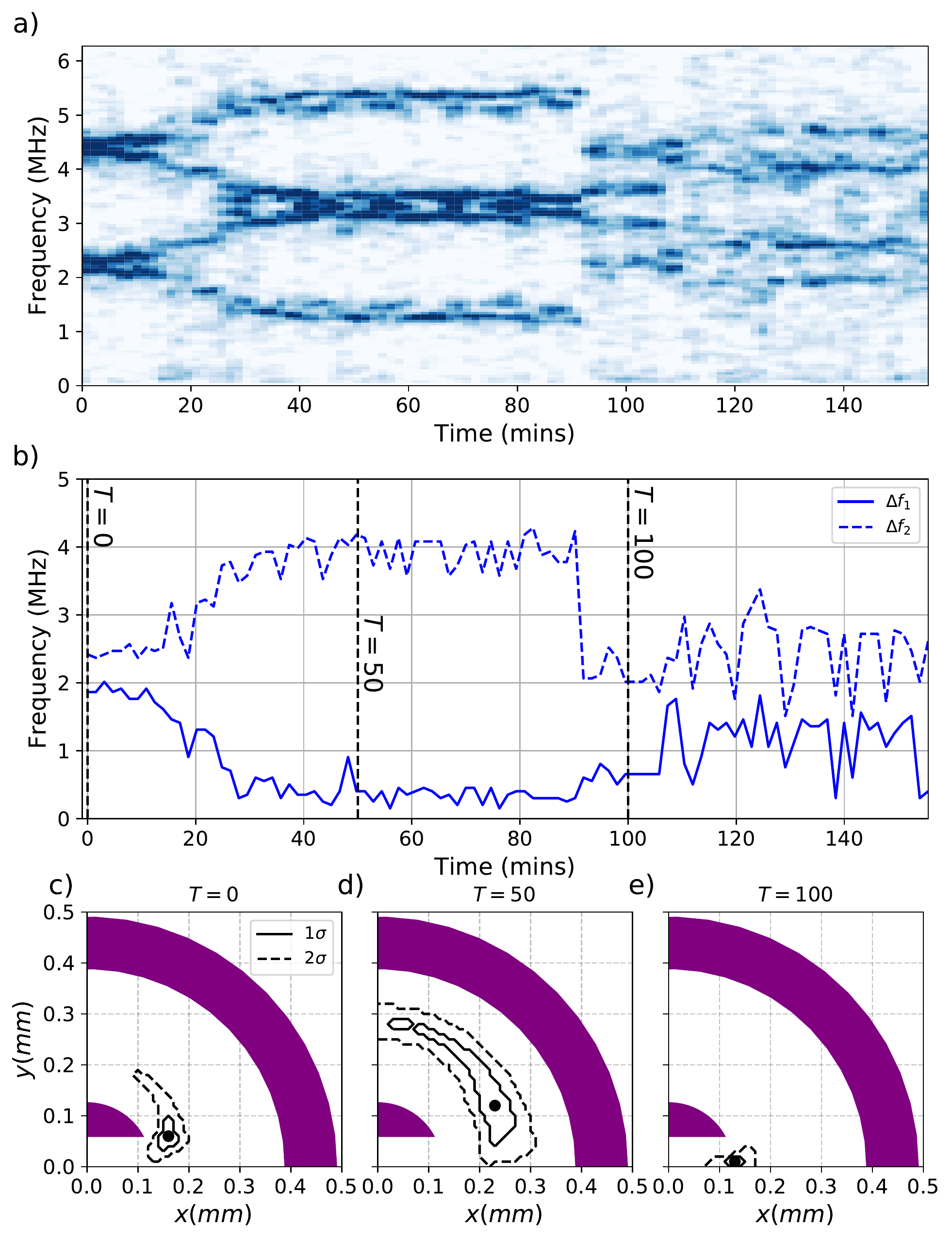}
\caption{Time series measurement of charge dispersion. (a) Fourier transform of repeated Ramsey oscillation experiments performed on the $\Sigma$-mode, over a period of 2.5 hours as shown in Fig.~\ref{figure:3}. (b) Tracking of inner frequency peak separation (solid line), $\Delta f_1$, and outer frequency peak separation (dashed line), $\Delta f_2$, from time series measurement shown in (a). Black vertical dashed lines highlight $T = 0$ mins, $T = 50$ mins and $T = 100$ mins. (c), (d), (e) Mapping of charge configuration to physical location of charge on substrate at highlighted times shown in (b). Uncertainty in position due to error in fitting of Fourier transform data, shown for $1\sigma$ (solid line) and $2\sigma$ (dashed line), with the most likely position indicated by the black dot. One quarter of the device is shown due to the four-fold symmetry of the system.}
\label{figure:4}
\end{figure}

Through repeated Ramsey interferometry of device A, monitored over the course of 10 hours, we find a maximum dispersion of $\epsilon_{10} / h= 4$ MHz ($\epsilon_{01} / h = 8$ MHz) for the $\Sigma$-mode ($\Delta$-mode). This is consistent with the order of magnitude of our predicted values of 4 MHz (4.1 MHz). The larger difference between $\epsilon_{10}$ and $\epsilon_{01}$ is due to asymmetries in the Josephson energies of the junctions, caused by fabrication imperfections. 

In device B, we perform Ramsey experiments on the more charge sensitive $\ket{01}- \ket{11}$ transition, and find no frequency beating up to a resolution of 10 kHz, demonstrating that this device design iteration has a suppressed sensitivity to charge noise, by at least a factor of 400 compared to device A. This measurement is shown in Appendix E.

However, this suppression of charge sensitivity in device B comes at a cost of lower mode anharmonicities, and state-dependent shifts, as shown in Table I. This reduces the maximum speed at which single-qubit gates, and entangling operations between modes can be completed. 

\section{SPATIALLY RESOLVED CHARGE DETECTION}

We now use device A to demonstrate a proof-of-concept detector of localised charge fluctuations. By repeating the Ramsey interferometry we can track the frequency dispersion $\Delta f_1$ and $\Delta f_2$, and using our tight-binding model, we are able to infer a time-series measurement of charge configuration of $n_{g\Sigma}$ and $n_{g\Delta}$. We find the expected spatial charge sensitivity of these sum and difference modes using an electrostatic simulation, shown in Fig.~\ref{figure:1} (d) and (e). As the induced gate-charge offsets $n_{g\Sigma}$ and $n_{g\Delta}$ exhibit different spatial sensitivities, we can deduce the position of a potential surface-charge on the device, using a biangulation method. For a given value of $n_{g\Sigma}$, we use the simulation data shown in Fig.~\ref{figure:1} (d) to identify an area where a charge ($1e$) would induce a gate-charge offset of that value. We repeat this for a corresponding value of $n_{g\Delta}$ with the simulation data shown in Fig.~\ref{figure:1} (e). The overlap of these two individually obtained areas allows us to identify the location of a charge ($1e$) on the surface of the substrate. However, the use of two modes produces an ambiguity in the quadrant in which any surface charge is located. Future devices could use three or more islands, and incorporate a symmetry breaking geometry, in order to more accurately triangulate surface charge position.

We perform a demonstrative tracking experiment using repeated Ramsey interferometry measurement over a 150 minute period, as shown in Fig.~\ref{figure:4}. We observe both slow frequency drifts corresponding to fluctuating charge configuration, as well as a singular large frequency jump, indicative of non-equilibrium charge dynamics \cite{Martinis2021}. We convert the measured frequency dispersion to give a spatial estimation of charge location at time T = 0, T = 50, and T = 100 minutes, indicated by the dashed vertical lines on Fig.~\ref{figure:4} (b).

From $T = 0$ to $T = 50$ minutes, we observe a slow drift in surface charge moving outwards away from the inner islands of the device. The uncertainty in position from $T = 50$ onwards is high, as the device is not sensitive to spatial fluctuations far from the two inner islands, shown by the lower gradient in the simulation of induced charge in Fig.~\ref{figure:1}. In this period we also observe a much clearer signal in the raw FFT data of Fig.~\ref{figure:4} (a), consistent with the charge configuration remaining stable throughout the individual Ramsey oscillation measurement.

From $T = 100$ onwards, we observe a shift in the surface charge distribution towards the two inner islands of the device. In this region, we have a much higher spatial resolution, shown by the larger gradient in simulation of induced charge from Fig.~\ref{figure:1}. As such, any small movement in surface charge distribution will cause a larger shift in charge configuration of the device. We observe this as a noisier measurement of frequencies shown in Fig.~\ref{figure:4} (b). This is also shown in the raw FFT data where the peaks are less clear in this region, consistent with the charge configuration changing during the course of the measurement. With an improved measurement rate, this detector could be used for observing spatial charge fluctuations with a resolution of less than $100$ $\mu m$.

%% file: conclusion.tex
In this work, we have investigated charge sensitivity in a two-mode transmon. Our results show observations of multiple charge parity configurations, and show agreement with a predictive theory for charge sensitivity using a tight binding approximation. In addition, we show this sensitivity can be suppressed for high coherence quantum computing applications.

We use a two-mode device in the charge-sensitive regime (device A) to demonstrate proof-of-principle spatially-resolved charge detection. Combining a similar device with high fidelity state readout in future would enable detection of parity jumps between three-superconducting islands, and determine how quasi-particle tunnelling events affect energy dissipation in multiple modes. If local control of static electric field were incorporated in a future device via gate electrodes, it would become possible to unambiguously translate frequency fluctuations to precise charge configurations, for precise detection of local differential charge-noise vs. global charge-noise. We hence propose such multi-mode charge-sensitive qubits as potentially powerful tools for ongoing investigations into sources of decoherence in superconducting quantum devices.

%% file: appendix.tex
\subsection{Circuit Quantisation}
The two-mode coaxial transmon has the Hamiltonian:
\begin{equation}
\hat{H} = 8E_{C_\Sigma}\hat{n}_\Sigma^2 + 8E_{C_\Delta}\hat{n}_\Delta^2 - 2E_J \cos{\frac{\hat{\varphi_\Sigma}}{2}} \cos{\frac{\hat{\varphi_\Delta}}{2}}
\tag{A1}
\label{equation:7}
\end{equation}
where $E_{C_\Sigma} = E_c + E_p/2 = e^2 / 2(C^*-C_m)$ and $E_{C_\Delta} = E_c - E_p/2 = e^2 / 2(C^*+C_m)$ are the charging energy of the sum and difference modes respectively, as described in the main text.
This shows independent charge variables $\hat{n}_\Sigma,~\hat{n}_\Delta$ with a coupled potential energy term $U(\varphi_\Sigma, \varphi_\Delta) = - 2E_J \cos{\frac{\varphi_\Sigma}{2}} \cos{\frac{\varphi_\Delta}{2}}$. 
We can predict the behaviour of this system in the transmon regime $E_p \ll E_c \ll E_J$, by expanding the cosine potential terms. To fourth order in $\varphi_\Sigma,~\varphi_\Delta$, we find:
\begin{equation}
\begin{aligned}
    \hat{H} = 8E_{C_\Sigma}\hat{n}_\Sigma^2 + 8E_{C_\Delta}\hat{n}_\Delta^2\\
    + 2E_J (1/8 \hat{\varphi}_\Sigma^2 + 1/8 \hat{\varphi}_\Delta^2 \\
    - 1/{24 * 4^4}\hat{\varphi}_\Sigma^4 - 1/{24 * 4^4}\hat{\varphi}_\Delta^4 \\
    - 1/{16} \hat{\varphi}_\Sigma^2 \hat{\varphi}_\Delta^2) 
\end{aligned}
\tag{A2}
\end{equation}
Assuming a weakly nonlinear oscillator-like behaviour, we take $\hat{\varphi}_{\Sigma (\Delta)}=\sqrt{\frac{2E_J}{E_{C_\Sigma(\Delta)}}}(\hat{a}_{\Sigma(\Delta)}^\dagger + \hat{a}_{\Sigma(\Delta)})$, and $\hat{n}_{\Sigma (\Delta)}=\frac{i}{2}\sqrt{\frac{E_{C_\Sigma(\Delta)}}{2E_J}}(\hat{a}_{\Sigma(\Delta)}^\dagger - \hat{a}_{\Sigma(\Delta)})$, where $\hat{a}_{\Sigma(\Delta)}^{(\dagger)}$ is the annihilation (creation) operator for the $\Sigma$-mode ($\Delta$-mode). Keeping only counter-rotating terms produces:
\begin{equation}
\begin{aligned}
    \hat{H}/\hbar &= \omega_\Sigma \hat{a}^\dagger_\Sigma \hat{a}_\Sigma + 
    \omega_\Delta \hat{a}^\dagger_\Delta \hat{a}_\Delta \\
    &-\eta_\Sigma/2 *  \hat{a}^\dagger_\Sigma{}^2 \hat{a}_\Sigma{}^2- \eta_\Delta/2 *  \hat{a}^{\dagger}_\Delta{}^2 \hat{a}_\Delta{}^2 \\
    &- 4\sqrt{\eta_\Sigma \eta_\Delta} * \hat{a}^\dagger_\Sigma \hat{a}_\Sigma \hat{a}^\dagger_\Delta \hat{a}_\Delta
\end{aligned}
\tag{A3}
\end{equation}
where $\eta_{\Sigma(\Delta)}$ is the anharmonicity of the $\Sigma$-mode ($\Delta$-mode). Notice that the state dependent shift term $4\sqrt{\eta_\Sigma \eta_\Delta} \equiv \chi_{\Sigma\Delta}$ can be greater than the individual mode anharmonicities $\eta_{\Sigma(\Delta)}$.

The full Hamiltonian of the system shown in Fig.~\ref{figure:1} (a) is given by $\hat{H}_{full} = \hat{H} + \hat{H}_{r}$, where $\hat{H}_{r}$ describes the coupling between each mode and the readout resonator, given by:

\begin{equation}
\begin{aligned}
    \hat{H}_{r}/\hbar &= \omega_r \hat{a}^\dagger_r \hat{a}_r +  g_{\Sigma(\Delta) r}(\hat{a}_{r}^\dagger + \hat{a}_{r})(\hat{a}_{\Sigma(\Delta)}^\dagger + \hat{a}_{\Sigma(\Delta)})
\end{aligned}
\tag{A4}
\end{equation}

where $\hat{a}^{(\dagger)}_r$ is the annihilation (creation) operator for the resonator mode, $\omega_r$ is the resonator mode frequency, and $g_{\Sigma(\Delta) r}$ is the coupling between the resonator mode and the $\Sigma$-mode ($\Delta$-mode). This weak coupling between each transmon mode and the resonator mode leads to a mode-state-dependent resonator frequency shift $\chi_{\Sigma(\Delta)r}$ as defined in ref. \cite{Rahamim2017}. This allows us to perform simultaneous dispersive readout of the multi-mode state of the device.

\subsection{Device Parameters}
\begin{figure}
\includegraphics[width=0.42\textwidth]{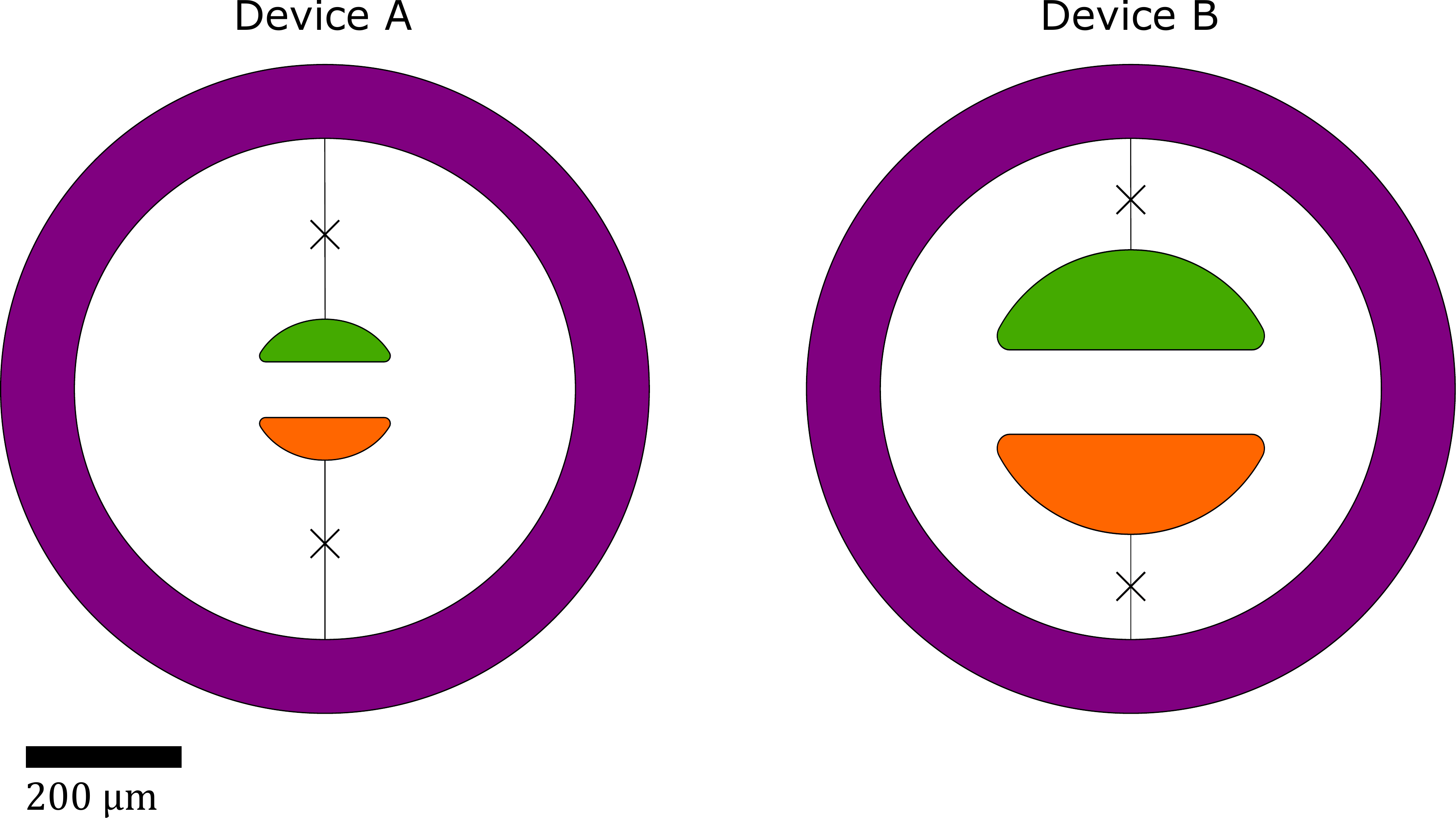}
\caption{Comparison of device designs. Device A (left) has an inner island radius of $125 ~\rm \mu m$, and inner island gap of $120 ~\rm \mu m$. Device B (right) has an inner island radius of $220 ~\rm \mu m$, and inner island gap of $125 ~\rm \mu m$. Both devices have the same outer island geometry, with inner radius of $389.5 ~\rm \mu m$, and outer radius of $489.5 ~\rm \mu m$.}
\label{figure:5}
\end{figure}

In Table~\ref{table:I} we show the measured parameters of the devices used in these experiments. Fig.~\ref{figure:5} shows the geometry of both the charge sensitive (A) and insensitive (B) two-mode coaxial transmons. 
\begin{table}[ht]
\centering
\caption{Device Parameters}
\begin{tabularx}{0.45\textwidth}{l r r}
\hline
&Device A&Device B \\
\hline
\textbf{LC Resonator} &&\\
Frequency $f_r$ [GHz] & 9.72 & 9.32\\
Linewidth $\kappa_r /2\pi$ [MHz] & 2.8 & 1.9\\
Dispersive Shift $2\chi_{\Delta r} /2\pi$ [MHz] & 3.9 & 2.1\\
Dispersive Shift $2\chi_{\Sigma r} /2\pi$ [MHz] & 4.9 & 1.7\\
&&\\
\textbf{$\Delta$ - Mode }&&\\
Transition Frequency $\omega_\Delta / 2\pi$ [GHz] & 5.51 & 4.58\\
Anharmonicity $\eta_\Delta / 2\pi$ [MHz] & 380 & 104\\
$T_1$  [$\mu s$] & 40.7 & 50.9\\
$T_{2}$ Echo [$\mu s$] & 33.8 & 24.8\\
&&\\
\textbf{$\Sigma$ - Mode} &&\\
Transition Frequency $\omega_\Sigma / 2\pi$ [GHz] & 6.71 & 5.74\\
Anharmonicity $\eta_\Sigma / 2\pi$ [MHz] & 340 & 144\\
$T_1$  [$\mu s$] & 13.5 & 26.5\\
$T_{2}$ Echo [$\mu s$] & 13.1 & 19.4\\
&&\\
Cross Kerr Shift $\chi_{\Sigma \Delta} / 2\pi$ [MHz] & 500 & 269\\
\hline
\end{tabularx}
\label{table:I}
\end{table}

\subsection{Wavefunctions}
\begin{figure}
\includegraphics[width=0.46\textwidth]{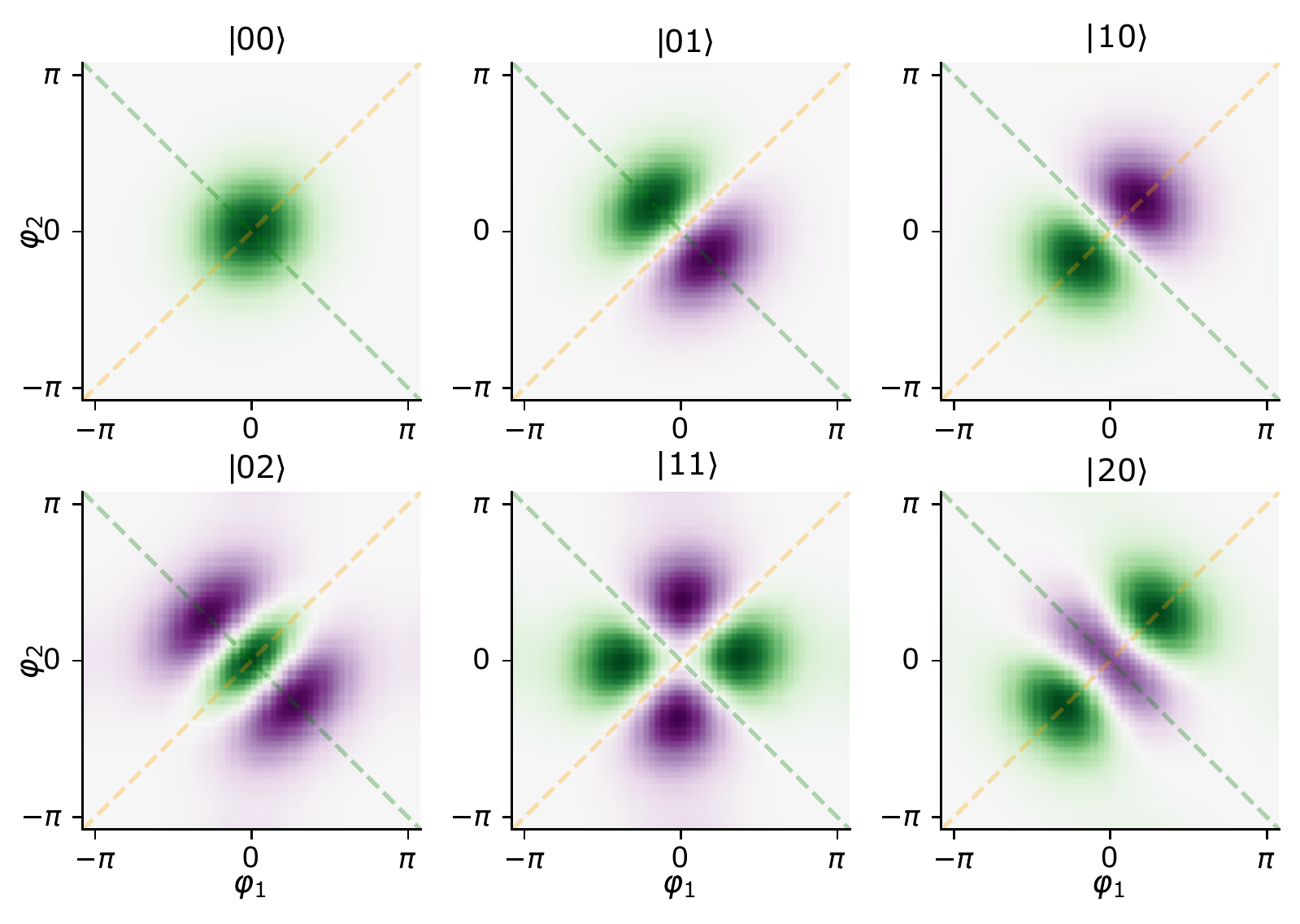}
\caption{Wavefunctions of the six lowest energy eigenstates of the two-mode coaxial transmon. Diagonal dashed lines indicate the summation ($(\varphi_1 + \varphi_2)$, orange) and difference ($(\varphi_1 - \varphi_2)$, green) phases. States labelled as $\ket{nm}$ where $n (m)$ corresponds to the number of excitations in the $\Sigma$-mode ($\Delta$-mode).}
\label{figure:6}
\end{figure}

Fig.~\ref{figure:6} shows the wavefunctions of the six lowest energy eigenstates of the two-mode coaxial transmon Hamiltonian in Eqn.~\ref{equation:7}. This Hamiltonian was constructed and solved in the charge basis and then transformed to the phase basis to obtain the wavefunctions plotted, using the package scqubits \cite{groszkowski2021scqubits}.

\subsection{Tight-Binding Model}
The Hamiltonian of Eqn.~\ref{equation:7} can be written as $\hat{H} = \hat{T} + \hat{U}$, where the kinetic term $\hat{T} = 8E_{C_\Sigma}\hat{n}_\Sigma^2 + 8E_{C_\Delta}\hat{n}_\Delta^2$, and the potential term can be rewritten as:
\begin{equation}
\begin{aligned}
    U(\varphi_\Sigma, \varphi_\Delta) &= - 2E_J \cos{\frac{\varphi_\Sigma}{2}} \cos{\frac{\varphi_\Delta}{2}} \\
    & =-2E_J\cos{\frac{\varphi_\Sigma}{2}} -2E_J\cos{\frac{\varphi_\Delta}{2}}\\ &                 -4E_J\sin^2{\frac{\varphi_\Sigma}{4}} \sin^2{\frac{\varphi_\Delta}{4}}\\
    &= U_0(\varphi_\Sigma, \varphi_\Delta) + U_1(\varphi_\Sigma, \varphi_\Delta)
\end{aligned}
\tag{A5}
\end{equation}
where $U_0 = -2E_J\cos{\frac{\varphi_\Sigma}{2}} -2E_J\cos{\frac{\varphi_\Delta}{2}}$, is the potential term for two uncoupled transmons, and $U_1 = -4E_J\sin^2{\frac{\varphi_\Sigma}{4}} \sin^2{\frac{\varphi_\Delta}{4}}$, is a perturbation coupling the two transmon modes together. The potential $U_0$ is $4\pi$ periodic in $\varphi_\Sigma$ and $\varphi_\Delta$, and the additional potential term $U_1$ introduces lattice sites at $(\varphi_\Sigma, \varphi_\Delta) = (\pm 2\pi, \pm 2\pi)$.

Given this lattice structure, we use Bloch's theorem to pick an approximate solution to the Schr\"{o}dinger equation $\hat{H}\ket{\psi} = E \ket{\psi}$ as:

\begin{equation}
\begin{aligned}
    \psi(\vec{\varphi}) &= \frac{1}{\sqrt{2}} (\psi^{(1)}(\vec{\varphi}) + \psi^{(2)}(\vec{\varphi})) \\
    &=\frac{1}{\sqrt{2}} e^{i\vec{k}.\vec{\varphi}}(u_k(\vec{\varphi}) + e^{i\vec{k}.\vec{a}}u_k(\vec{\varphi} - \vec{a}))
\end{aligned}
\tag{A6}
\end{equation}
where $\vec{k}$ is the wave vector of the wavefunction $\psi$, $\vec{\varphi}$ is the position vector given by $(\varphi_\Sigma, \varphi_\Delta)$,  $\vec{a}$ denotes the lattice site introduced by the additional potential $U_1$, and $u_k$ is a $4\pi$ periodic function. Due to the symmetry of the system, we only need to consider the lattice sites located at $\vec{\varphi} = (0, 0)$ and $\vec{\varphi} = (2\pi, 2\pi)$.

We calculate the energies of the system with this approximate wavefunction using:

\begin{equation}
\begin{aligned}
    E(k) = \frac{\bra{\psi_k}\hat{H}\ket{\psi_k}}{\bra{\psi_k}\ket{\psi_k}}
\end{aligned}
\tag{A7}
\end{equation}

leading to the dispersion relation:
\begin{equation}
\begin{aligned}
    E(n_{g\Sigma}, n_{g\Delta}) \approx E_0 + \cos{\pi n_{g\Sigma}}\cos{\pi n_{g\Delta}}(\gamma - \alpha \beta)
\end{aligned}
\tag{A8}
\end{equation}

where the wave vector $\vec{k} = (n_{g\Sigma}/2, n_{g\Delta}/2)$, and the quantities $\alpha$, $\gamma$ and $\beta$ are the tight binding coefficients given by:
\begin{equation}
\begin{aligned}
    \alpha = 4 \int_{0}^{2\pi} u_k(\vec{\varphi})u_k(\vec{\varphi}-2\pi) \,d^2\vec{\varphi} 
\end{aligned}
\tag{A9}
\end{equation}

\begin{equation}
\begin{aligned}
    \beta = -16E_J \int_{0}^{2\pi} |u_k(\vec{\varphi})|^2 \sin^2{\frac{\varphi_\Sigma}{4}} \sin^2{\frac{\varphi_\Delta}{4}}\,d^2\vec{\varphi} 
\end{aligned}
\tag{A10}
\end{equation}

\begin{equation}
\begin{aligned}
    \gamma = -16E_J \int_{0}^{2\pi} u_k(\vec{\varphi})u_k(\vec{\varphi}-2\pi) \sin^2{\frac{\varphi_\Sigma}{4}} \sin^2{\frac{\varphi_\Delta}{4}}\,d^2\vec{\varphi} 
\end{aligned}
\tag{A11}
\end{equation}

The dominant term here is $\gamma$, which describes the bond energy between wavefunctions at adjacent lattice sites, also known as the two center integral. The $\alpha$ and $\beta$ terms describe the overlap integral between wavefunctions on adjacent lattice sites, and the energy shift due to the potential on neighbouring lattice sites respectively. These two terms are small compared to $\gamma$ and so can be neglected, leading to the approximate form of the maximum charge dispersion $\epsilon_{mn}/4 \approx \gamma$.

This form of the charge dispersion $\epsilon_{mn}$ can either be calculated numerically, using the wavefunctions shown in Appendix C, or analytically using a semi analytical wavefunction approach to obtain the functional form of $u_k$ \cite{Catelani2011}. Using this approach, we arrive at the analytical form for the maximum charge dispersion shown in Eqn.~\ref{equation:5}.

\subsection{Ramsey Experiments On The Charge Insensitive Two-Mode Transmon}
In order to demonstrate a suppression of offset charge sensitivity in device B, we perform Ramsey interferometry measurements on the $\ket{01} - \ket{11}$ transition. From our predictive model, we expect a maximum charge dispersion of $\epsilon_{11}/h \approx 10$ kHz.

\begin{figure}
\includegraphics[width=0.48\textwidth]{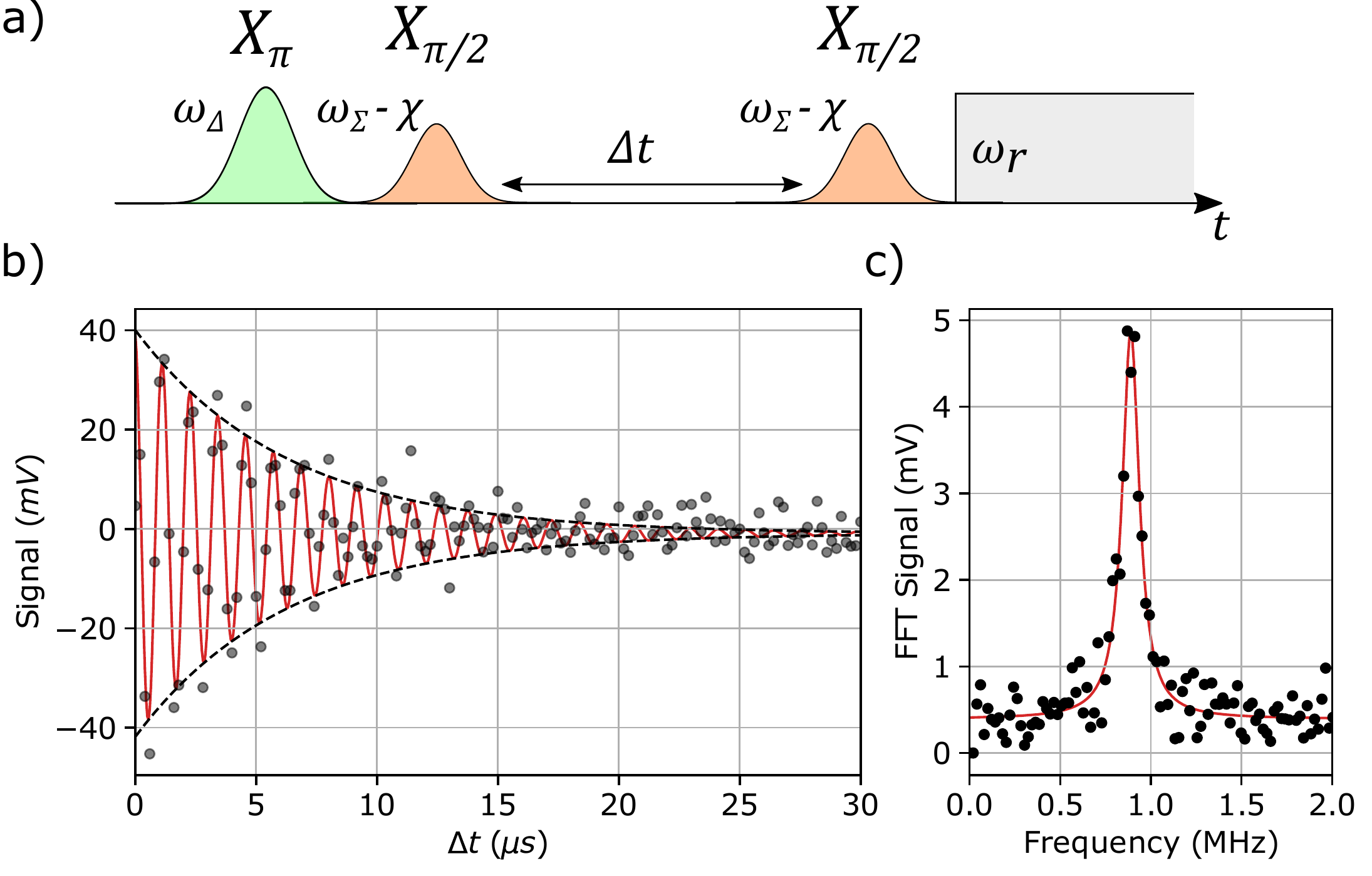}
\caption{Ramsey interferometry on the $\ket{01} - \ket{11}$ transition of the charge insensitive two-mode transmon (device B). (a) Pulse sequence for the Ramsey interferometry measurement. (b)(c) Ramsey oscillation and Fourier transform of the measurement, with data points in black, and fits in solid lines. The data in (b) is fitted to a decaying oscillation, and the dashed lines show an exponential decay envelope with the same decay constant. The data in (c) is fitted to a single Lorentzian peak.}
\label{figure:7}
\end{figure}

In order to measure this dispersion, we first prepare the system in the $\ket{01}$ state using a $X_\pi$ pulse, at a frequency $\omega_\Delta$. We then place the system in a superposition using a $X_{\pi/2}$ pulse, at a frequency $\omega_\Sigma - \chi$. The system is then left to idle for time $\Delta t$, before a second $X_{\pi/2}$ pulse is applied, as shown in the pulse scheme, shown in Fig.~\ref{figure:7} (a). As with the previous Ramsey interferometry measurements, the experiment is sampled many times, and so we expect to average over all possible parity configuration. 

In Fig.~\ref{figure:7}, we show the measured Ramsey oscillation, and FFT, for this transition. In contrast to the measurements performed on the charge sensitive device A, shown in Fig.~\ref{figure:3}, the oscillation is comprised of a single frequency component. This is demonstrated in the FFT of the decaying oscillation, shown in Fig.~\ref{figure:7} (c), where we are unable to distinguish a further frequency component, to a resolution of 10 kHz. This demonstrates this device design iteration has a suppressed sensitivity to charge noise.

